\documentstyle[12pt,psfig,epsfig,axodraw]{article}
\advance\textheight by \topskip
\begin{document}
\tolerance=100000
\thispagestyle{empty}
\setcounter{page}{0}

\def\cO#1{{\cal{O}}\left(#1\right)}
\newcommand{\be}{\begin{equation}}
\newcommand{\ee}{\end{equation}}
\newcommand{\br}{\begin{eqnarray}}
\newcommand{\er}{\end{eqnarray}}
\newcommand{\ba}{\begin{array}}
\newcommand{\ea}{\end{array}}
\newcommand{\bi}{\begin{itemize}}
\newcommand{\ei}{\end{itemize}}
\newcommand{\bn}{\begin{enumerate}}
\newcommand{\en}{\end{enumerate}}
\newcommand{\bc}{\begin{center}}
\newcommand{\ec}{\end{center}}
\newcommand{\ul}{\underline}
\newcommand{\ol}{\overline}
\newcommand{\ra}{\rightarrow}
\newcommand{\sm}{${\cal {SM}}$}
\newcommand{\as}{\alpha_s}
\newcommand{\aem}{\alpha_{em}}
\newcommand{\ycut}{y_{\mathrm{cut}}}
\newcommand{\susy}{{{SUSY}}}
\newcommand{\Dir}{\kern -6.4pt\Big{/}}
\newcommand{\Dirin}{\kern -10.4pt\Big{/}\kern 4.4pt}
\newcommand{\DDir}{\kern -10.6pt\Big{/}}
\newcommand{\DGir}{\kern -6.0pt\Big{/}}
\def\Ecm{\ifmmode{E_{\mathrm{cm}}}\else{$E_{\mathrm{cm}}$}\fi}
\def\gluino{\ifmmode{\mathaccent"7E g}\else{$\mathaccent"7E g$}\fi}
\def\photino{\ifmmode{\mathaccent"7E \gamma}\else{$\mathaccent"7E \gamma$}\fi}
\def\mgluino{\ifmmode{m_{\mathaccent"7E g}}
             \else{$m_{\mathaccent"7E g}$}\fi}
\def\taugluino{\ifmmode{\tau_{\mathaccent"7E g}}
             \else{$\tau_{\mathaccent"7E g}$}\fi}
\def\mphotino{\ifmmode{m_{\mathaccent"7E \gamma}}
             \else{$m_{\mathaccent"7E \gamma}$}\fi}
\def\ML{\ifmmode{{\mathaccent"7E M}_L}
             \else{${\mathaccent"7E M}_L$}\fi}
\def\MR{\ifmmode{{\mathaccent"7E M}_R}
             \else{${\mathaccent"7E M}_R$}\fi}
\def\lsim{\buildrel{\scriptscriptstyle <}\over{\scriptscriptstyle\sim}}
\def\gsim{\buildrel{\scriptscriptstyle >}\over{\scriptscriptstyle\sim}}
\def\jp #1 #2 #3 {{J.~Phys.} {#1} (#2) #3}
\def\pl #1 #2 #3 {{Phys.~Lett.} {#1} (#2) #3}
\def\np #1 #2 #3 {{Nucl.~Phys.} {#1} (#2) #3}
\def\zp #1 #2 #3 {{Z.~Phys.} {#1} (#2) #3}
\def\pr #1 #2 #3 {{Phys.~Rev.} {#1} (#2) #3}
\def\prep #1 #2 #3 {{Phys.~Rep.} {#1} (#2) #3}
\def\prl #1 #2 #3 {{Phys.~Rev.~Lett.} {#1} (#2) #3}
\def\mpl #1 #2 #3 {{Mod.~Phys.~Lett.} {#1} (#2) #3}
\def\rmp #1 #2 #3 {{Rev. Mod. Phys.} {#1} (#2) #3}
\def\sjnp #1 #2 #3 {{Sov. J. Nucl. Phys.} {#1} (#2) #3}
\def\cpc #1 #2 #3 {{Comp. Phys. Comm.} {#1} (#2) #3}
\def\xx #1 #2 #3 {{#1}, (#2) #3}
\def\NP(#1,#2,#3){Nucl.\ Phys.\ \issue(#1,#2,#3)}
\def\PL(#1,#2,#3){Phys.\ Lett.\ \issue(#1,#2,#3)}
\def\PRD(#1,#2,#3){Phys.\ Rev.\ D \issue(#1,#2,#3)}
\def\preprint{{preprint}}
\def\Ord{\lower .7ex\hbox{$\;\stackrel{\textstyle <}{\sim}\;$}}
\def\OOrd{\lower .7ex\hbox{$\;\stackrel{\textstyle >}{\sim}\;$}}
\def\MCH {$\tilde\chi_1^+$}
\def \CH{{\tilde\chi}^{\pm}}
\def \LSP{\tilde\chi_1^0}
\def \SNU{\tilde{\nu}}
\def \BARSNU{\tilde{\bar{\nu}}}
\def \MLSP{m_{{\tilde\chi_1}^0}}
\def \MCH{m_{{\tilde\chi}^{\pm}}}
\def \MCHMIN {\MCH^{min}}
\def \ET{\not\!\!{E_T}}
\def \LL{\tilde{l}_L}
\def \LR{\tilde{l}_R}
\def \MLL{m_{\tilde{l}_L}}
\def \MLR{m_{\tilde{l}_R}}
\def \MSNU{m_{\tilde{\nu}}}
\def \PROCESS{e^+e^- \rightarrow \tilde{\chi}^+ \tilde{\chi}^- \gamma}
\def \PI{{\pi^{\pm}}}
\def \DM{{\Delta{m}}}
\newcommand{\bQ}{\overline{Q}}
\newcommand{\ad}{\dot{\alpha }}
\newcommand{\bd}{\dot{\beta }}
\newcommand{\dd}{\dot{\delta }}
\def \CH{{\tilde\chi}^{\pm}}
\def \N0{{\tilde\chi^0}}
\def \MCH{m_{{\tilde\chi}_1^{\pm}}}
\def \LSP{\tilde\chi_1^0}
\def \MUL{m_{\tilde{u}_L}}
\def \MUR{m_{\tilde{u}_R}}
\def \MDL{m_{\tilde{d}_L}}
\def \MDR{m_{\tilde{d}_R}}
\def \MSNU{m_{\tilde{\nu}}}
\def \MLL{m_{\tilde{l}_L}}
\def \MLR{m_{\tilde{l}_R}}
\def \mhf{m_{1/2}}
\def \MST{m_{\tilde t_1}}
\def \RPVC{\lambda'}
\def\tth{\tilde{t}\tilde{t}h}
\def\qqh{\tilde{q}_i \tilde{q}_i h}
\def\t1{\tilde t_1}
\def \pt{p{\!\!\!/}_T}  
\def\lapp{\mathrel{\rlap{\raise.5ex\hbox{$<$}}
                    {\lower.5ex\hbox{$\sim$}}}}
\def\gapp{\mathrel{\rlap{\raise.5ex\hbox{$>$}}
                    {\lower.5ex\hbox{$\sim$}}}}
\begin{flushright}
TIFR/HEP/04-02
\end{flushright}
\begin{center}
{\Large \bf
Constraining R-parity violating couplings using dimuon data at Tevatron Run-II.
}\\[1.00
cm]
\end{center}
\begin{center}
{\large Subhendu Chakrabarti, Monoranjan Guchait  
{and} N. K. Mondal}\\[0.3 cm]
{\it Department of High Energy Physics\\
Tata Institute of 
Fundamental Research\\ 
Homi Bhabha Road, Bombay-400005, India.}
\end{center}

\vspace{2.cm}

\begin{abstract}
The dimuon plus dijet signal is analyzed in the top squark pair production 
at Tevatron Run-II experiment and the total event rate is compared with
the existing dimuon data. This comparison rules out top
squark mass upto 188(104)~GeV for the branching fraction 100\%(50\%) of 
top squark decay into the muon plus quark via lepton number violating coupling.
Interpretation of this limit in the framework of R-parity violating(RPV) 
SUSY model puts limit on relevant RPV coupling for a given top
squark mass and other supersymmetric model parameters. 
If $\MST \lsim 180$~GeV we found that the RPV couplings are roughly restricted 
to be within $\sim 10^{-4}$ which is at the same ballpark value 
obtained from the neutrino data. The limits are very stringent for a 
scenario where top squarks appear to be the next lightest supersymmetric 
particles.        

{\noindent\normalsize 
}
\end{abstract}
\vspace{2cm}
\newpage

\label{sec_intro}
The Minimal Supersymmetric Standard Model(MSSM)~\cite{susy} is a theoretically 
well motivated and also a very strong candidate for beyond standard model(SM) 
physics. Although,so far, there is no experimental evidence of 
Supersymmetric(SUSY) theory, nevertheless, in all current and future 
high energy collider based experiments, looking for SUSY is one of the 
very high priority programs. The MSSM contains
all SM particles in addition to their corresponding
SUSY partners and at least two higgs doublets which are the ingredient to 
switch on the electroweak symmetry breaking mechanism to generate masses 
of all physical particles. Lack of evidence in degeneracy of masses among 
particles and corresponding sparticles
implies that SUSY is not a exact symmetry, it has to be broken.  
In MSSM, a mixing occurs among different 
chiral states of sfermions($\tilde f$), superpartners of fermions($f$). 
Since the extent of mixing is proportional to the corresponding 
fermion mass, $m_f$, naturally, the mixing between the third generation 
of left and 
right handed sfermions, $\tilde f_L, \tilde f_R$ becomes more stronger 
than the case of other two 
generations of sfermions~\cite{stopmix}. As a consequence, there is a large 
splitting between the mass
eigen states 
$\tilde t_1$ and $\tilde t_2$(assume 
$\MST \leq m_{\tilde t_2}$)
of top squarks(SUSY partners of top quarks). Moreover, because of large 
Yukawa coupling,
the soft SUSY masses $m_{\tilde t_L}, m_{\tilde t_R}$ receive a significant 
corrections via the renormalization group equation~\cite{RGstop} which 
results more splitting between
the masses of $\t1$ and $\tilde t_2$ states. Incidentally, it may happen 
that for a 
certain region of
SUSY parameter space the lighter state of top squark, $\t1$, turns out to be
the next lightest SUSY particle(NLSP), where the lightest 
neutralino, $\LSP$, is assumed to be the lightest SUSY 
particles(LSP).              

The decay pattern of $\t1$ is phenomenologically very interesting. In the 
R-parity conserving(RPC) model a 
scenario where $\MST$ is heavier than the mass of the lighter 
chargino $m_{\CH_1}$, the charged current decay mode of $\t1$ via a b quark 
and $\CH_1$ 
\br
\t1 \ra b  \CH_1 
\label{eq:chdk}
\er
dominates. Otherwise, $\t1$
mainly decays via the loop induced flavor changing neutral current 
decay mode~\cite{hikasa},
\br
\t1 \ra c  \N0_1 
\label{eq:loopdk} 
\er
and as well as via the four body decay 
mode into a b quark, LSP and two massless fermions,
\br
\t1 \ra b \N0_1 f \bar f'
\label{eq:4bodydk} 
\er
The competition between these two decay channels in absence of charged 
current decay mode, Eq.~\ref{eq:chdk} mainly 
controlled 
by SUSY parameter space~\cite{abdel}. Along with all these decay channels, 
in the framework of R-parity violating(RPV) SUSY model, $\t1$ can have 
also other 
decay channels via lepton number violating couplings of class 
$\lambda'_{i3j}$, 
\br
\t1 \ra \ell + q 
\label{eq:rpvdk}
\er
assuming baryon number violating interactions are forbidden.
Here $i$($j$) stand for the lepton(quark) family index.
Of course, the search strategy of top squark in hadron colliders   
is decided by the decay pattern of $\t1$. In RPC SUSY model, 
the detailed study
of top squark searches has already been presented in 
Ref.~\cite{tata,matchev,yaan} 
where as the discovery potential of $\t1$ for RPV SUSY model are discussed
in recent studies~\cite{spdas}. 

In RPV SUSY model, neglecting the masses of leptons and quarks,
the decay width
of $\t1$ into a lepton and a quark solely depends, 
for a given  RPV $\lambda'_{i3j}$ coupling, on the mass of $\t1$ state 
and $\cos\theta_{\tilde t}$, 
where $\theta_{\tilde t}$ is the mixing angle between two chiral states 
$\tilde t_L$ and $\tilde t_R$. As a consequence of this decay channel, 
in this model, the pair production of top squark is signaled by di-lepton 
plus di-jet without 
missing energy in the final states
{\footnote { In case of $\tau$ lepton some amount of missing energy 
will appear 
because of the presence of $\nu_\tau$ from $\tau$ decay.}. 
For a given luminosity, the total event rate, of course, depends on the 
branching ratio, 
$\epsilon_{br}^{\ell}$=BR($\t1 \ra \ell + q$) which is given by, 
\br
Br(\t1 \ra \ell + q) = \frac { \Gamma_{R{\!\!\!/}} (\t1 \ra \ell + q) }
{\Gamma_{R{\!\!\!/}}(\t1 \ra \ell + q) + \Gamma_R(\t1 \ra all)}
\label{eq:brt1}
\er
Here $\Gamma_{R{\!\!\!/}} (\t1 \ra \ell + q)$ stands for the total decay width 
of $\t1$ in the RPV channel where as  $\Gamma_R(\t1 \ra all)$ presents
the total decay width of $\t1$ for all possible kinematically allowed 
channels in RPC model. Needless 
to say that for a given $\MST$, $\epsilon_{br}^{\ell}$ is very 
sensitive to $\lambda'_{i3j}$ and  $\cos\theta_{\tilde t}$ and as well as to  
the total decay width of $\t1$ in all accessible RPC decay channels of $\t1$. 
However, the total decay width of $\t1$ in all RPC channels depends on  
SUSY parameters, mainly sensitive to $M_2$ - the SU(2) 
gaugino mass{\footnote{Assuming the 
gaugino mass relation $M_1 \simeq M_2/2$}, $\mu$ - the higgsino mass
parameter and $\tan\beta$, the ratio of two vacuum expectation values 
required to 
generate masses of particles in the model. The 
pattern of $\epsilon_{br}^{\ell}$ has been investigated in detail for a wide 
range of SUSY parameter space~\cite{spdas}. It is found that the RPV decay 
mode, Eq.\ref{eq:rpvdk}, is very competitive and dominates over the other 
two decay channels, Eq.\ref{eq:loopdk} which is loop suppressed
and Eq.~\ref{eq:4bodydk} which is
suppressed because of its four body final state.   
Interestingly, for a substantial
region of SUSY parameter space, this RPV decay mode dominates over these 
two decay modes even for a smaller value of RPV couplings
$\lambda'_{i3j}({\sim 10^{-4}})$~\cite{spdas}. However,
this is not the case when $\t1$ is allowed to decay via charged current 
decay mode, Eq.\ref{eq:chdk}, in the scenario $\MST \gsim \MCH$ i.e when 
$\t1$ is not the NLSP. In this scenario, the RPV decay channel requires 
large value of $\lambda'_{i3j}$ to make it competitive with the charged 
current decay mode~\cite{spdas}.

We can argue that, if the RPV SUSY model be a viable model, then
the signature of this model may be found through the top squark pair 
production which has comparatively larger cross section as $\t1$ is 
likely to be the lightest colored sparticle via its decay 
channel Eq.\ref{eq:rpvdk}.   
The search prospect of $\t1$ at Run-II in Tevatron 
experiment in the 
framework of RPV model has been discussed in detail in Ref.~\cite{spdas} for 
the class of $\lambda'_{i3j}$ RPV coupling. 
The discovery potential of $\t1$ are thoroughly discussed in the 
dilepton plus
dijet channel in a model independent way. The range of $\lambda'_{i3j}$ 
which can be probed for a given luminosity is also presented~\cite{spdas}. 
The present study is devoted to 
investigate the signal of $\t1$ state in the di-muon plus dijet 
($\mu\mu$ + jj) channel because of the  
decay $\t1 \ra \mu + q$ via the lepton number violating RPV 
coupling $\lambda'_{23j}$. 
Eventually, a comparison is made between our predicted event rates 
with the existing preliminary data which was analyzed to study the signal 
of second generation of Leptoquark searches in D0 detector at Tevatron Run-II 
experiment~\cite{d0collab} with $\sqrt{s}=$1.96 TeV.
It is worth to mention here that in our earlier study, we analyzed the 
di-electron plus dijet final state from top
squark pair production and its subsequent decay, $\t1 \ra e + q$, 
via RPV coupling $\lambda'_{13j}$~\cite{subhendu} and compared the predicted
event rates with the existing dielectron data in D0 detector at Run-I. The 
di-electron plus di-jet data was reported in the context 
of first generation of Leptoquark searches of which final state event topology 
is same as the final state containing di-electrons plus di-jets.
We concluded from that study that $\MST \gsim $220~GeV can be ruled out 
for 100\% decay of 
$\t1$ in the channel,Eq.~\ref{eq:rpvdk}. Moreover,
in the framework of RPV SUSY model we excluded certain region in the  
$\MST -\lambda'_{13j}$ plane for a given SUSY parameter 
space~\cite{subhendu}.    

It is interesting to note that for a long time RPV SUSY models has been
known as a viable option which can provide models of neutrino mass
~\cite{rpvneumass}. These models have attracted special attention after the
neutrino data confirms that the neutrinos are not massless~\cite{neutrino}.
Clearly, the parameters which participate in the process of neutrino mass
generation in a given model can be constrained using neutrino data
and obviously it will be model dependent prediction~\cite{rpvneut,abada}.   
For example, for a certain class of models where $\lambda'_{i33}$
lepton number violating couplings are required to generate neutrino masses
are restricted to be within the range of $\sim 10^{-3} - 10^{-4}$ depending on 
the magnitude of soft breaking masses in RPC SUSY theory~\cite{abada}. 
Certainly, these bounds are more stronger than the previous bounds prior 
to neutrino data~\cite{han}. Notably, our previous 
analysis in the $ee+jj$ channel predicts upper bounds on the same set of 
RPV couplings which are at the same ballpark~\cite{subhendu}.
At Tevatron Run-II, the implications of these relevant RPV 
couplings has been studied in detail very systematically~\cite{spdas}. 
We notice that it is quite possible to find the signal of RPV SUSY if the 
RPV couplings are in the vicinity of these predicted bounds.                   
This interesting observation motivates us to further extend our study in 
the $\mu\mu + jj$ channel in RPV SUSY model and similarly as 
before~\cite{subhendu}, examine the value of 
RPV coupling $\lambda'_{23j}$ allowed by the existing dimuon 
data~\cite{d0collab}. 
In the past, using Tevatron data there was a attempt to constrain 
squark and gluino masses~\cite{dp} and RPV couplings~\cite{asesh} 
in the framework of RPV SUSY model.
In the next section we describe our analysis and discuss our results with 
a summary at the end.

At Tevatron top squark pair production takes place via 
$q \bar q, g g \ra \t1 \t1^*$ and the magnitude of cross section depends 
mainly on $\MST$~\cite{qcdcs}. The QCD 
correction enhances
the cross section by $\sim$30\% roughly, although this correction depends 
on the choice of QCD scales $Q^2$~\cite{spira} which we set to $Q^2=\hat {s}$. 
In our calculation we used CTEQ5L parametrization for incoming parton flux. 
As explained earlier, focusing our signal to $\mu\mu$+jj
final state which arises due to $\t1$ decay, 
$\t1 \ra \mu + q$ via $\lambda'_{23j}$ 
RPV coupling we intend to compare total event
rates of this final state with the 
existing Run II di-muon result~\cite{d0collab}. 
The event generator {\tt PYTHIA}~\cite{pythia} 
has been used to 
generate events from $\t1$
pair production and then forcing $\t1$ to decay in the 
$\t1 \ra \mu + q$ channel.
The hadronisation effects with the initial and final state radiation 
has been considered during event generation. We have
used {\tt PYCELL}~\cite{pythia} for jet formation. 
Note that we have not performed any detector simulation in our analysis.
We have used {\tt PYTHIA} mainly to find the geometrical and kinematic
selection efficiencies 
for the set of cuts defined in ~\cite{d0collab}. Eventually, we obtain the 
signal 
cross section by multiplying selection efficiency along with all other 
detection
efficiencies due to triggering, muon isolation and identification, 
jet reconstruction and tracking efficiencies 
as quoted in Ref.~\cite{d0collab}.

In our analysis we have applied the same set of kinematic cuts
~\cite{d0collab} on final state 
particles which are optimized mainly to suppress the SM 
backgrounds, 
particularly from $t \bar t$, 
$WW$ pair production and Drell-Yan process. The selection cuts are as 
following: 
(i) muons are selected if, 
$p_T^\mu >$ 15 GeV,$|\eta|<$1.9 and dimuon invariant mass
$M_{\mu\mu}>$ 60 GeV, muon isolation is confirmed by demanding $E_{0.4} - 
E_{0.1}< $ 2.5~GeV where $E_{0.4(0.1)}$ is  sum of the energy of the particles 
contained in a cone of size 0.4(0.1) in $\eta$-$\phi$ space around muons.
Note that we smear the muon momenta as done in Ref.~\cite{d0collab}.    
(ii) jets are required to have $E_T^j >$ 25 GeV, $|\eta_j|<$2.4 and two or 
more jets are accepted with $\Delta R=$0.4, $\Delta R=\sqrt{\Delta\eta^2 + 
\Delta\phi^2}$. The efficiency for jet reconstruction with $E_T^j >$ 30 GeV 
is given to be greater than $96 \pm 3$\%~\cite{d0collab}; 
(iii) Dimuon invariant mass is demanded to be $M_{\mu\mu}>$110 GeV; and 
(iv) $S_T > $ 280 GeV where $S_T = {\sum}_{i=\mu\mu jj}E_T^i$.    

Cuts (i) and (ii) are basically the preselection requirements to 
select events. Selection cut
(iii) is mostly to eliminate the background, mainly, 
from $Z \rightarrow \mu \mu$ where the events are distributed
around the Z-peak, and finally, selection criteria (iv) on scalar 
sum of visible transverse energy takes care of rest of the background 
cross sections. The number of signal events are:
\br
n_{sig} = \sigma_{\t1\t1}.\epsilon_{br}^{\mu}.
\epsilon_{ac}.\tilde L 
\label{eq:sigcs} 
\er
$\sigma_{\t1\t1}$
is the top squark pair production cross section for $\sqrt{s}$=1.96 TeV, 
$\epsilon_{ac}$ is the acceptance efficiency including tracking efficiency
(0.78$\pm$.007) and jet reconstruction efficiency(0.96$\pm$0.03). 
In the data, tracking efficiency is obtained with a sample of triggered
di-muon events from $Z \ra \mu\mu$. It is found that there is a dependence 
of tracking effciency on muon directions. Because of this $\eta$
dependence tracking effciciency needed to be parameterised\cite{d0collab}.
In our analysis we have used the same parametrization for tracking 
efficiency, which is about $\sim$ 65\%. The effective 
luminosity $\tilde L=L.c$, where c is the correction factor due to the trigger 
efficiency, muon identification and isolation 
efficiency~\cite{d0collab}. 
With the correction factor,
the effective luminosity turns out to be $\tilde L=$81.8 $\pm$ 9.1 pb$^{-1}$ 
where as the measured luminosity is 104pb$^{-1}$. In the calculation 
of cross section limits we have 
used this effective luminosity along with its error. In Table 1, we present, 
for various choices of $\MST$,
accepted efficiencies ($\epsilon_{ac}$) which is basically kinematic selection 
efficiencies computed by {\tt PYTHIA} after applying the selection cuts as 
described above, folded with jet reconstruction and  
tracking efficiencies. It is expected that the efficiencies increase with 
$\MST$ as for higher values of $\MST$ muons and jets become more and 
more harder.

Clearly, from the knowledge of $\sigma_{\t1\t1}, \epsilon_{ac}$
( see Table. 1) and $\tilde L$, the number of $\mu\mu$+jj events can 
be predicted using Eq.~\ref{eq:sigcs}. 
It is reported~\cite{d0collab} that there is only one 
event in the data and the estimated number of background events are 
to be 1.59$\pm$0.47; 
the uncertainties in background estimation 
are purely due to the systematic and statistical errors. Armed with this 
information 
we attempt to 
find the cross section limits using the Bayesian method assuming flat 
prior cross section distribution for different 
choices of $\MST$. In Fig.1, we display the limits of top squark pair 
production
cross section by solid lines for various $\MST$ values and for two choices 
of $\epsilon_{br}^{\mu}$          
=0.5 and 1. The dashed line represents the predicted theoretical Born level 
cross section multiplied by K-factor(=1.3)~\cite{spira}
{\footnote {Although the value of 
K-factor depends on the choice of QCD scales, but for the sake of simplicity
we assumed it to be fixed.}}.
This figure clearly shows that the top squark mass upto  
$\MST \gsim$ 188(104)~GeV is ruled out for $\epsilon_{br}^{\mu}$=1.0(0.5). 
We want to emphasize here that the limits of $\MST$ as shown in Fig.1 is 
completely
model independent.
The limiting values of top squark pair 
production cross sections as shown in Fig.1 and Eq.~\ref{eq:sigcs} 
enable us to predict upper limits
of $\epsilon_{br}^{\mu}$ for each value
of $\MST$. In Fig.2, for various choices of $\MST$, we present the
maximum values of $\epsilon_{br}^{\mu}$ at 95\% C.L, which are allowed
by existing dimuon data. 
As for example, for $\MST$=100 GeV, the limiting value of top squark pair 
production cross section from data 
forbids $\epsilon_{br}^{\mu} \gsim $48\%, otherwise
top squark signal in this channel could be observed. 
The shaded region in Fig.2 is excluded at 95\% C.L and certainly,
it is a model independent prediction. Clearly, for $\MST \gsim $180~GeV
there is no limit on $\epsilon_{br}^{\mu}$ since event rate is negligible for
smaller values of cross sections in this mass range.  
It is already mentioned that in a given model the branching 
ratio  $\epsilon_{br}^{\mu}$  is controlled by the model 
parameters. Recall that the decay width of the channel, 
Eq.~\ref{eq:rpvdk},  
is very sensitive to $\lambda'_{23j}$ RPV coupling and mixing 
angle $\theta_{\tilde t}$. In addition to that
the branching ratio,$\epsilon_{br}^{\mu}$, is also indirectly controlled
by, mainly by 
$M_2,\mu$ and $\tan\beta$, which actually determine the total 
decay width 
of $\t1$ in all RPC decay channels.
Hence, for a given SUSY parameter space the total decay width of $\t1$ 
in all RPC channels is fixed. Consulting eq.\ref{eq:brt1} and for a fixed 
value of $\cos\theta_{\tilde t}$ one can obtain the limiting values of
RPV coupling $\lambda'_{23j}$ from the upper limits of   
$\epsilon_{br}^{\mu}$. Following this strategy, for
the purpose of illustration, in Fig. 3, we display the excluded region in 
the $\MST-\lambda'_{23j}$ plane for two sets of values
of $\tan\beta$=5(Fig.3a) and 30(Fig.3b) setting the SUSY parameters
to  $M_2=130$~GeV, $\mu=500~$GeV. 
The other SUSY parameters which are involved, but not very sensitive to 
our results are shown in the figure caption. 
This selected set of SUSY parameters determine 
the value of $m_{\CH_1}=$122(125)~GeV; $m_{\N0_1}=$63(65)~GeV.
In each figure, we present excluded region for two nearly extreme values
of $\cos\theta_{\tilde t}$=0.95 and 0.02. 
These figures indicate that the 
RPV coupling $\lambda'_{23j}$ is bounded roughly by 
$\sim 10^{-3} - 10^{-4}$ for the region of parameter space  
where $\MST \lsim m_{\tilde\chi_1^\pm}$ and $\tan\beta$=5(see Fig.3a).  
For higher $\tan\beta$ case(=30), this limit turns out to be smaller 
approximately by one order of magnitude (see Fig.3b). However, the limit 
becomes weaker for the scenario where $\MST \gsim m_{\tilde\chi_1^\pm}$, 
because of the fact that in this region, 
the decay mode, $\t1 \ra b + \CH_1$ opens up and dilutes the dimuon plus 
dijet event rate. The enhancement of RPV decay width with the increase of
$\cos\theta_{\tilde t}$ value results more excluded region than the
region for low value of $\cos\theta_{\tilde t}$. On the other hand
higher value of $\tan\beta$(=30) results enhanced loop decay width,
Eq.~\ref{eq:loopdk}, costing branching ratio in the RPV decay 
channel, which eventually exclude comparatively narrow region, as shown in 
Fig.3b. It is interesting to note that limits obtained from the
present analysis in the $\t1$ NLSP scenario for $ \MST \lsim $180~GeV
are comparable to the limits obtained from the neutrino data. 
Needless to say that this conclusion is very much SUSY parameter space
dependent and holds for that region of parameter space where
top squark is light($\lsim 180$~GeV) and appear to be the NLSP.

In summary, we have computed the top squark pair production signal cross 
section 
in the di-muon plus dijet channel and compared the 
event rates with the existing data in D0 detector at the 
Tevatron Run-II experiment. 
This type of signal occurs in the context
of RPV SUSY model in the presence of lepton number violating coupling 
$\lambda'_{23j}$ which is assumed to be the dominant one.
Our analysis rules out the top squark mass, $m_{\tilde t_1} \gsim$ 188(104)~GeV
in a model independent way
for the branching fraction of top squark in the muon plus jet 
channel, $\epsilon_{br}^{\mu}$=1(0.5)(see Fig.1). The top squark pair 
production cross section limits obtained from data for each $\MST$ 
restrict the value of $\epsilon_{br}^{\mu}$(see Fig.2). In the framework 
of RPV SUSY model, the upper limits of $\epsilon_{br}^{\mu}$ can be 
translated to the upper limits of $\lambda'_{23j}$ coupling for a given 
set of SUSY parameters. We found that our predicted limits on 
$\lambda'_{23j}$ are very close to the limits obtained from neutrino data 
for a certain region of SUSY parameter space where $\MST$ behaves like 
NLSP and $\MST \lsim$ 180~GeV.
  
\section*{Acknowledgment}
The authors are thankful to Tim Christiansen 
for useful discussion and inputs concerning the analysis.

\newpage
\hspace{5 cm}
\vspace{1 cm}
\begin{center}
\begin{tabular}{|c|c|}
\hline
$m_{\tilde t_1}$(GeV) & Detection efficiency(\%) \\

&$\mu^{+}\mu^{-}$    \\
\hline
100 & 1.52 \\
120& 3.4 \\
140& 6.67 \\
160& 10.4 \\
180& 13.8\\
200& 15.7  \\
220& 17.2 \\
\hline
\end{tabular}
\end{center}

Table 1: Di-muon plus di-jet detection efficiencies for various  $\MST$.

\begin{center}
\begin{figure}[htb]
\hspace*{1.2cm}
\includegraphics[width=5.in, height=3.in]{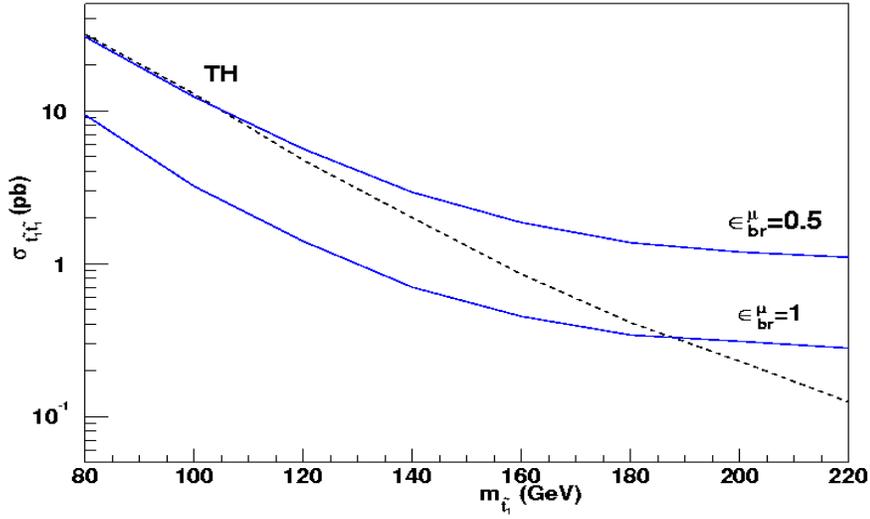}
\vspace{1.cm}
\caption
{The top squark pair production cross section limits using data 
at 95\% C.L. for 
$\epsilon^{\mu}_{br}=$1 and 0.5(solid lines). The dashed line 
represents the theoretical prediction. 
}
\label{fig_cslimit}
\end{figure}
\end{center}
\begin{figure}[htb]
\vspace{-3cm}
\hspace*{1.4cm}
\includegraphics[width=5.in, height=3.in]{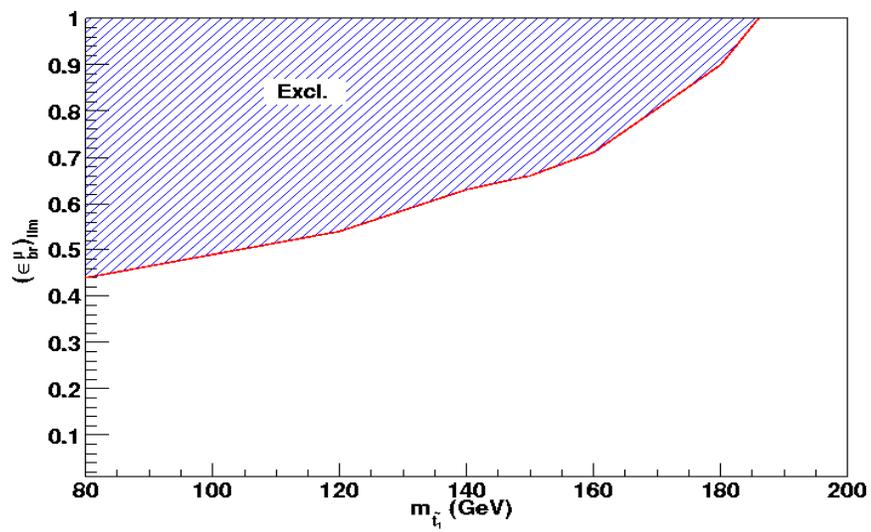}
\caption
{ The excluded region(shaded area) by dimuon data at 95\% C.L.  
}
\label{fig_brlimit}
\end{figure}

\begin{figure}[htb]
\vspace{-1.3cm}
\includegraphics[width=5in, height=3.in]{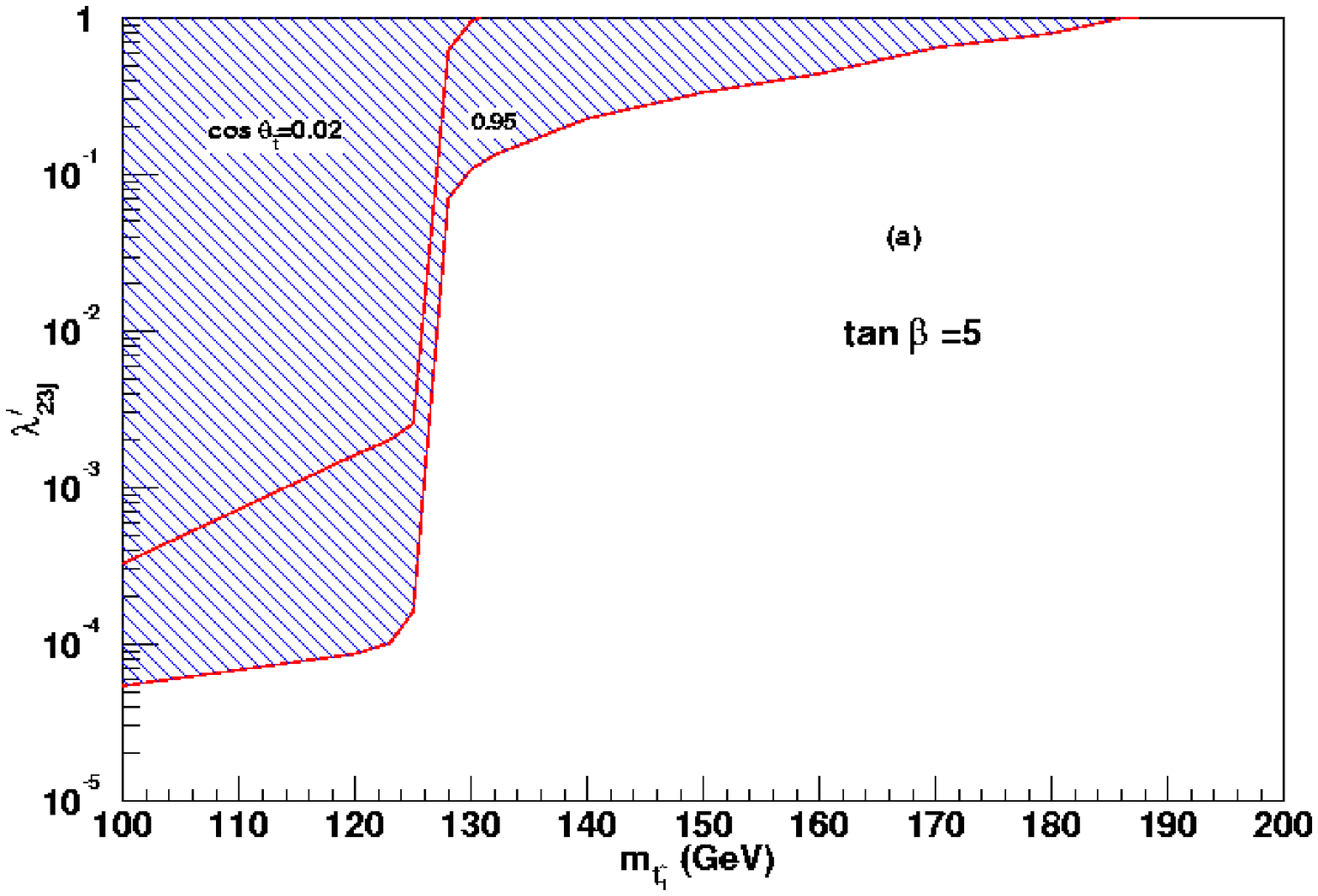} 
\hspace{2cm}\includegraphics[width=5in, height=3.in]{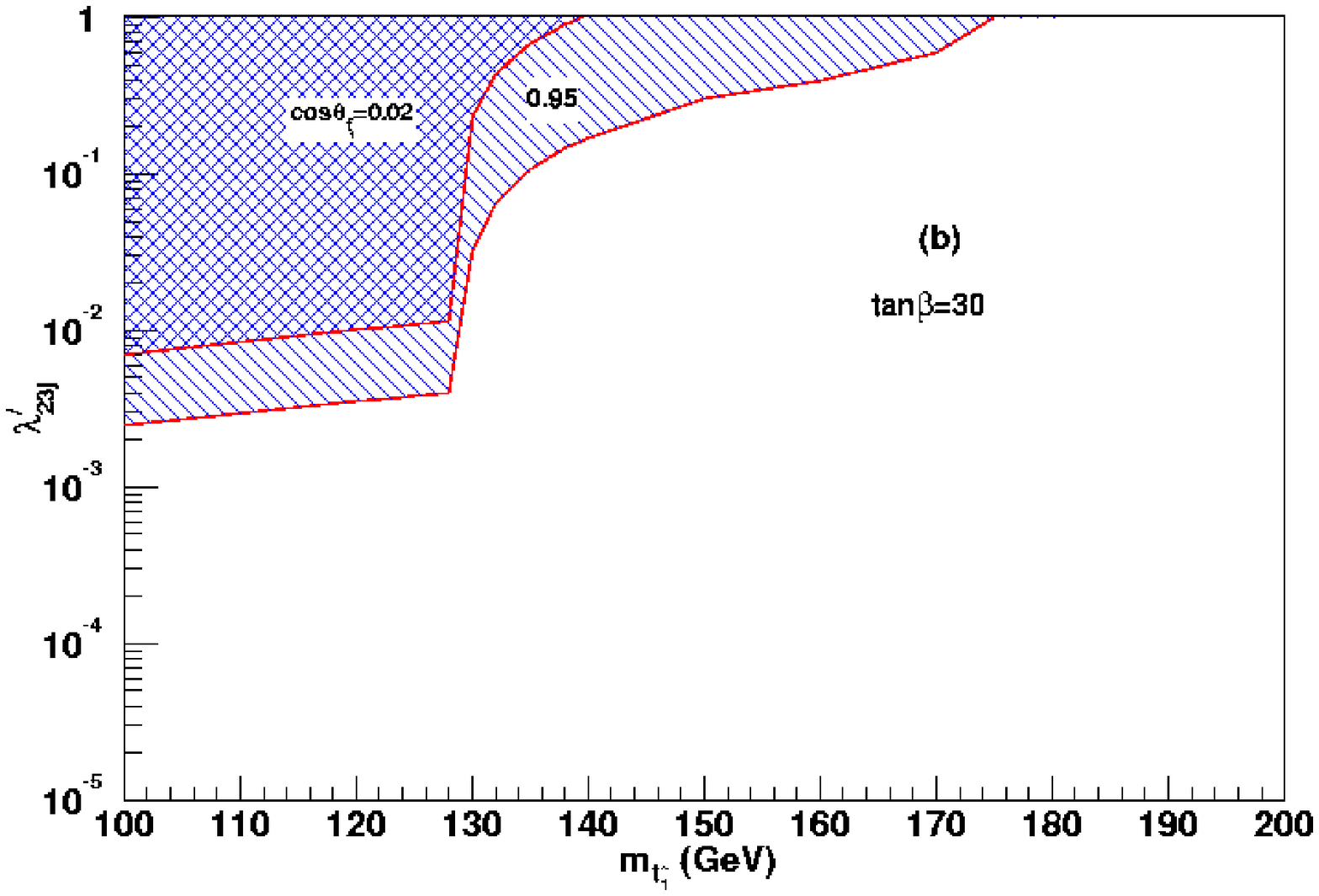}
\caption
{The excluded region(hatched) by di-muon data
at 95\% C.L. The SUSY parameters are: $M_2=$130 GeV, $\mu$=500 GeV,
$m_{\tilde q}=$300 GeV, $m_{\tilde\ell}=$200 and A-terms=200 GeV.
}
\label{fig_lamlimita}
\end{figure}


\begin{thebibliography}{99}

\bibitem{susy} 
For reviews of the Supersymmetry, see e.g.
H. E. Haber and G. Kane, Phys. Rep. 117,75(1985); M. Drees and
S. Martin, CLTP Report (1995) and {\tt hep-ph/9504324}


\bibitem{stopmix} J. Ellis and S. Rudaz, Phys. Lett. B128,248(1983);
M. Drees and K. Hikasa, Phys. Lett. B252,127(1990).

\bibitem{RGstop}
S.Abel et al., Report of the ``SUGRA working group for 
``RUN-II at the Tevatron'', hep-ph/0003154 and references therein.


\bibitem{hikasa}  K.I. Hikasa and M. Kobayashi, Phys. Rev. D36,724(1987).

\bibitem{abdel} C. Boehm, A. Djouadi and Y. Mambrini, Phys. Rev. D61
,095006(2000). A. Djouadi and Y. Mambrini, Phys. Rev. D63, 115005(2001).

\bibitem{tata} H. Baer, M. Drees, R. M. Godbole and X. Tata, Phys. Rev.
D44,725(1991); H. Baer, J. Sender and X. Tata, {\it ibid} D50,4517(1994).

\bibitem{matchev}R. Demina, J. Lykken, K. Matchev and A. Nomerotski, Phys. Rev.
D62,035011(2000); M. Carena, D. Choudhury, R. A. Diaz, H. E. Logan
and C. E. M. Wagner, Phys. Rev. D66,115010(2002).

\bibitem{yaan} A. Djouadi, M. Guchait and Y. Mambrini,
Phys. Rev. D64, 095014(2001). S. P. Das, A. Datta  and M.Guchait, 
Phys. Rev. D65, 095006(2002).

\bibitem{spdas} Siba Prasad Das, Amitava Datta and 
M. Guchait,{\tt hep-ph/0309168}.

\bibitem{d0collab} Emmanuelle Perez,Talk presented at Lepton Photon '03,
Fermilab; Cuts used obtained fron D$\emptyset$ through private 
communication. 

\bibitem{subhendu} Subhendu Chakrabarti, M. Guchait and N. K. Mondal, 
Phys. Rev. D68,015005(2003) ({\tt hep-ph/03012148}).

\bibitem{rpvneumass} C. S. Aulakh and R. N.Mohapatra, Phys. Lett. B119,
136(1982); L. Hall and M. Suzuki, Nucl. Phys. B231,419(1984);
J. Ellis {\it et al.}, Phys. Lett. B150,142(1985);
G. Ross and J. Valle, Phys. Lett B151,375(1985); S. Dawson, Nucl.Phys. 
B261, 297(1985).

\bibitem{neutrino} Y.Fukuda {\it et. al}, Super-Kamiokande Collaboration, 
Phys. Rev. Lett. 81, 1562(1998); Q. R. 
Ahmed {\it et. al.}, SNO collaboration, Phys. Rev. Lett. 89, 011301(2002); 
{\it ibid} 89, 011302(2002); K. Eguchi {\it et .al}, KamLAND collaboration,
Phys. Rev. Lett. 90,021802(2003). 

\bibitem{rpvneut} M. Nowakowski, A. Pilaftsis, Nucl. Phys. B461,19(1996);
A. Y. Smirnov, F. Vissani, Nucl. Phys. B460,37(1996);
J. Valle, {\tt hep-ph/9712277};
B. Mukhopadhyaya, S. Roy, F. Vissani, Phys. Lett. B443,191(1998);
S Rakshit , G Bhattacharyya , A Raychaudhuri Phys. Rev. D59,091701(1999);
S.Y. Choi, E. J. Chun, S. K. Kang, J.S. Lee, Phys. ReV. D60,075002(1999);
A. S. Joshipura, S. Vempati, Phys. Rev. D60,111303(1999);
Y. Koide, A. Ghosal, {\tt hep-ph/0203113}; R Adhikari, A Sil, A Raychaudhuri
Eur.Phys.J.C25,125(2002); F. Borzumati, J. S. Lee,{\tt hep-ph/0207184}; 
M. A. Diaz, H. Hirsch, W. Porod, J.C. Romao and J. W. F. Valle, Phys.
Rev. D68, 013009(2003).M. Hirsch, W.Porod, Phys. Rev. D68,115007(2003);
G. Bhattacharya,{\tt hep-ph/0305330} and references therein.

\bibitem{abada} 
M. Drees, S. Pakvasa, X. Tata, T. ter Veldhuis, Phys. ReV. D57,5335(1998);
A. Abada, M. Losada, Nucl.Phys.B585, 45 (2000).

\bibitem{han}V. Barger, G. F. Giudice and T. Han,
Phys. Rev. D40, 2987 (1989);  G. Bhattacharyya, {\tt hep-ph/9709395};
B. C. Allanach, A. Dedes and H. K. Dreiner, Phys. Rev. D60,075014(1999).


\bibitem{dp}D.~P.~Roy, Phys. Lett. B283,270(1992).

\bibitem{asesh} Aseshkrishna Datta, Biswarup Mukhopadhyaya,
Phys. Rev. Lett 85,248(2000).

\bibitem{qcdcs} G. Kane and J. P. Leveille, Phys. Lett. B112, 227 (1982) ;
P. R. Harrison and C.H. Llewellyn-Smith, Nucl. Phys. B213, 223 (1983) ;
E. Reya and D. P. Roy, Phys. Rev. D32, 645(1985) ;
S. Dawson, E. Eichten and C. Quigg, {\it ibid } 31, 1581 (1985);
H. Baer and X. Tata, Phys. Lett. B160, 159 (1985).

\bibitem{spira} W. Beenakker, M. Kramer, T. Plehn, M. Spira and P. M.
Zerwas, Nucl. Phys. B515, 3 (1998).

\bibitem{pythia} T. Sjostrand, P. Eden, C. Friberg, L. Lonnblad, G. Miu, 
S. Mrenna and  E. Norrbin, Computer Physics Commun. 135, 238 (2001). 

\end{thebibliography}
\end{document}